\documentclass[10pt]{article}
\usepackage{epsf}

\usepackage{epsfig,graphics}
\usepackage{graphicx}
\usepackage{epsfig}
\usepackage{subfigure}
\usepackage{tabularx} 
\usepackage{rotate}	
\usepackage{slashed}

\usepackage{amsmath}
\usepackage{amsfonts}
\usepackage{amssymb}
\usepackage{graphicx}
\usepackage{xcolor}
\usepackage{cite}

\usepackage{fancyhdr}
\usepackage{hyperref}

\newcommand{\bmat}{\left(\begin{array}}
	\newcommand{\emat}{\end{array}\right)}

\def\yzero{\smash{\hbox{$y\kern-4pt\raise1pt\hbox{${}^\circ$}$}}}

\def\beq{\begin{equation}}
	\def\eeq{\end{equation}}
\def\beqa{\begin{eqnarray}}
	\def\eeqa{\end{eqnarray}}

\def\-{\hphantom{-}}

\def\s2{\frac{1}{\sqrt2}}

\def\beq{\begin{equation}}
	\def\eeq{\end{equation}}
\def\beqa{\begin{eqnarray}}
	\def\eeqa{\end{eqnarray}}

\def\IF{\relax{\rm I\kern-.18em F}}
\def\II{\relax{\rm I\kern-.18em I}}
\def\IP{\relax{\rm I\kern-.18em P}}
\def\IC{\relax\hbox{\kern.25em$\inbar\kern-.3em{\rm C}$}}
\def\IR{\relax{\rm I\kern-.18em R}}

\def\Dsl{\,\raise.15ex\hbox{/}\mkern-13.5mu D} 
\def\IZ{Z\kern-.4em  Z}

\def\lt{\tilde \Lambda}

\def\BH{\text{BH}}
\def\Mt{M_{\text{tower}}}
\def\LambdaIR{\Lambda_{\text{IR}}}
\def\Mtt{\tilde{M}_{\text{tower}}}

\catcode`\@=11   
\newdimen\@rotdimen
\newbox\@rotbox  

\def\@vspec#1{\special{ps:#1}}
\def\@rotstart#1{\@vspec{gsave currentpoint currentpoint translate
		#1 neg exch neg exch translate}}
\def\@rotfinish{\@vspec{currentpoint grestore moveto}}
\def\@rotr#1{\@rotdimen=\ht#1\advance\@rotdimen by\dp#1%
	\hbox to\@rotdimen{\hskip\ht#1\vbox to\wd#1{\@rotstart{90 rotate}%
			\box#1\vss}\hss}\@rotfinish}
\def\@rotl#1{\@rotdimen=\ht#1\advance\@rotdimen by\dp#1%
	\hbox to\@rotdimen{\vbox to\wd#1{\vskip\wd#1\@rotstart{270 rotate}%
			\box#1\vss}\hss}\@rotfinish}%
\def\@rotu#1{\@rotdimen=\ht#1\advance\@rotdimen by\dp#1%
	\hbox to\wd#1{\hskip\wd#1\vbox to\@rotdimen{\vskip\@rotdimen
			\@rotstart{-1 dup scale}\box#1\vss}\hss}\@rotfinish}%
\def\@rotf#1{\hbox to\wd#1{\hskip\wd#1\@rotstart{-1 1 scale}%
		\box#1\hss}\@rotfinish}%
\def\rotate{\@ifnextchar[{\@rotate}{\@rotate[l]}}
\def\@rotate[#1]#2{\setbox\@rotbox=\hbox{#2}\@nameuse{@rot#1}\@rotbox}

\catcode`\@=12

\topmargin
-1.5cm
\textwidth
15.5cm
\textheight
23.5cm
\oddsidemargin
0.7cm
\evensidemargin
0.7cm

\begin{document}

	\makeatletter
	\@addtoreset{equation}{section}
	\makeatother
	\renewcommand{\theequation}{\thesection.\arabic{equation}}
	\pagestyle{empty}
	\vspace{-0.2cm}
	\rightline{ IFT-UAM/CSIC-21-152}
	\vspace{1.2cm}
	\begin{center}
		
		
		\LARGE{ IR/UV Mixing, Towers of Species \\
			and  Swampland Conjectures
		}
		\\[12mm]
		\large{ A. Castellano$^1$, A. Herr\'aez$^2$ and L.E. Ib\'a\~nez$^1$
			\\[12mm]}
		\small{
			$^1$ {\em Departamento de F\'{\i}sica Te\'orica
				and Instituto de F\'{\i}sica Te\'orica UAM/CSIC,\\[-0.3em]
				Universidad Aut\'onoma de Madrid,
				Cantoblanco, 28049 Madrid, Spain}  \\[0pt]
			$^4$  {\em Institut de Physique Th\'eorique, Universit\'e Paris Saclay, CEA, CNRS\\ [-0.3em]
				Orme des Merisiers, 91191 Gif-sur-Yvette CEDEX, France} 
			\\[5mm]
		}
		\small{\bf Abstract} \\[6mm]
	\end{center}
	\begin{center}
		\begin{minipage}[h]{15.22cm}
			By applying the  Covariant Entropy Bound  (CEB) to an EFT in a  box of size  $1/\Lambda_{\text{IR}}$ one  obtains that the UV and IR cut-offs of the EFT are necessarily correlated.
			We argue that in a theory of Quantum Gravity (QG) one should identify the UV cutoff with the `species scale', 
			and give a general algorithm to calculate it in the case of multiple towers becoming light.
			One then obtains  an upper bound on the characteristic mass scale of the tower in terms of  the IR cut-off, given by
			$M_{\text{tower}}\lesssim (\Lambda_{\text{IR}})^{2\alpha_D}$ in Planck units, with $\alpha_D=(D-2+p)/2p(D-1)$,
			where  $p$ depends on the density of states. 
			Identifying the IR cut-off with a (non-vanishing) curvature in
			AdS one reproduces the statement of the AdS Distance Conjecture (ADC), also giving an explicit lower bound for the $\alpha$  exponent.
			In particular, we find that the CEB implies $\alpha \geq1/2$ in any dimension if there is a single KK tower, both in AdS and dS vacua.
			However values $\alpha <1/2$  are allowed if the particle tower is multiple or has a string component.  We also consider  the CKN constraint 
			coming from avoiding gravitational collapse which further requires in general $\alpha \geq 1/D$ for the lightest tower.
			We analyse the case of the DGKT-CFI class of Type IIA orientifold models
			and show  it has both particle and string towers below the species scale,  so that a careful analysis  of how the ADC is defined is needed. We find that this class of
			models obey but do not saturate the CEB.  The UV/IR constraints found  apply to  both AdS and dS vacua.  We comment on possible applications of these ideas to the
			dS Swampland conjecture as well as to the observed dS phase of the universe.
		\end{minipage}
	\end{center}
	\newpage
	\setcounter{page}{1}
	\pagestyle{plain}
	\renewcommand{\thefootnote}{\arabic{footnote}}
	\setcounter{footnote}{0}
	

	
	\tableofcontents
	
	\section{Introduction}
	\label{sec:intro}
	
	Traditionally when trying to combine a standard D-dimensional local field theory with Quantum Gravity (QG) one parameterizes the effects  of gravity in terms of higher dimensional operators suppressed by powers of the D-dimensional Planck mass $M_D$ 
	\beq 
	\mathcal{L}_{\text{EFT,QG}} = \mathcal{L}_{\text{EFT}}+ \sum_{n=D}^{\infty} \frac{\mathcal{O}_n}{M_D^{n-D}} \ .
	\eeq
	where $\mathcal{L}_{\text{EFT}}$ refers to the field theory Lagrangian ignoring QG effects. This is in agreement with the Wilsonian understanding of effective field theories (EFT) in which there is a natural separation of scales and the UV physics has no effect on the IR physics other than the said suppressed operators. However, there are reasons to believe that this approach actually ceases to be correct in the presence of QG. Thus e.g. in String Theory duality symmetries map the  light spectrum of the theory to the supermassive spectrum. Independently of String Theory, black hole physics also seems to indicate intricate connections between the UV and the IR of whatever matter theory we are considering in the presence of gravity. The precise relation between the UV and the IR physics in QG is still to be elucidated. Notice that the possible connection between IR and UV cut-offs $\Lambda_{\text{IR}}, \Lambda_{\text{UV}}$ is not just academic since e.g.  these quantities appear in the computation of loop corrections in any EFT, like the Standard Model (SM)
	\cite{CKN,BD,CK}. A simple parameterization for such IR/UV connection would be given by an expression of the form
	\beq 
	\Lambda_{\text{UV}}\ \lesssim \ \Lambda_{\text{IR}}^\delta \ M_D^{1-\delta} \ ,
	\label{general}
	\eeq
	with $\delta<1$ a (possibly model-dependent) constant. Note that in the limit $M_D\rightarrow \infty $ the constraint becomes trivial, as one would expect if we want such a connection to arise as a genuine QG constraint.  In this sense (\ref{general}) has the form of a {\it Swampland}  condition \cite{swampland,WGC,distance} (see \cite{review,Irene,Alvaro} for reviews). But for finite Planck mass such an expression would tell us that the IR and UV scales are somehow correlated. Naively, this expression looks too simple to be correct but, as we will argue, it may be motivated by holographic arguments and we will show that it is indeed obeyed for large classes of String Theory vacua. The fact that such a correlation may be motivated from holography has been considered in the past, see e.g. \cite{CKN,BD,CK,Thomas,Bramante,Riotto,Seo}. 
	In the present work we argue that in a theory of QG the appropriate UV cut-off should be identified with the {\it species scale} \cite{Dvali}, so that such correlations 
	imply the presence of towers  of states below $\Lambda_{\text{UV}}$. This identification requires a careful treatment of the generic case in which multiple towers 
	with possibly different densities (e.g. multiple KK and/or string towers) are present. We give an algorithm to compute the species scale in this general set-up, also identifying 
	which towers appear below the UV cut-off. We find 
	\beq
	M_{\text{tower}} \ \lesssim \ \Lambda_{\text{IR}}^{2\alpha_D}\ M_D^{1-2\alpha_D}
	\eeq
	in D-dimensions, with $\alpha_D$ a constant given in eq. \eqref{primcota} below. For the case of multiple towers, $M_{\text{tower}}$ should be identified with a geometric mean of the scales of the towers of particles,
	weighted by a density of species parameter, $p$.
	As described below, for those values  of $\alpha_D$ the Covariant Entropy Bound (CEB) can be saturated.
	However we note that, as pointed out in \cite{CKN,BD} it turns out that before the theory reaches the {\it species} cut-off there is a lower cut-off $\lt< \Lambda_{\text{UV}}$ of physical relevance above which 
	the EFT is naively expected to break down due to gravitational collapse. So one expects that only the EFT in a region below that cut-off is fully reliable. One of the conditions one obtains is that  $\Mt \sim \Lambda_{\text{IR}}^{2\alpha }$ with
	$\alpha \geq 1/D$.
	
	In an AdS vacuum, by identifying the IR cut-off with the curvature radius one recovers the AdS Distance Conjecture (ADC) of \cite{ADC} with an explicit bound for its exponent $\alpha\geq \alpha_D$.
	In particular one finds $\alpha\geq 1/2$  for the case of a \textit{single} KK tower of states (for any $D$). Thus, all string vacua analysed so far are in agreement with this Covariant Entropy Bound.
	However values $1/D\leq \alpha \leq1/2$ are possible in the case of multiple KK and/or string towers. We argue that the Type IIA CY$_3$ orientifold examples of \cite{DGKT,CFI} belong to this class
	and obey, although do not saturate the Covariant Entropy Bound.
	All these observations still apply if we identify the IR cut-off with the vacuum scale in a dS vacuum.
	We briefly comment on the application of these ideas to the dS Conjecture \cite{Obied:2018sgi,K,dS}  and find some constraints in its rate parameter $c$.
	We also apply it to the observed cosmological constant, assuming it corresponds to a dS background in the universe,  and describe how towers of massless particles 
	would appear in this context.
	We also expect similar arguments to be relevant in the context of the Distance Conjecture \cite{distance}.\footnote{Note that entropy arguments have been also used to justify the asymptotic behavior of the scalar potential in the formulation of the dS Swampland Conjecture 
		\cite{dS} (see also \cite{review}, and \cite{Geng:2019bnn, Geng:2019zsx} for more applications to dS), and in connection with the Distance and Weak Gravity Conjectures in \cite{Hamada:2021yxy}.}

	\section{The Species Scale and Multiple Towers}	
	\label{sec:towers}

	Before delving into the (unexpected) consequences of having IR/UV mixing, it is important to clarify first what we mean here by the UV cut-off $\Lambda_{\text{UV}}$. This should be identified with the scale at which quantum gravitational effects can no longer be ignored, i.e. the so-called {\it species scale}. In a generic D-dimensional EFT weakly coupled to Einstein gravity this scale is given (up to numerical factors) by \cite{Dvali}
	\beq
	\Lambda_{\text{UV}}\ \simeq \ \frac {M_D}{N_{\text{species}}^{\frac{1}{(D-2)}}} \ ,
	\label{species}
	\eeq
	where $N_{\text{species}}$ is the number of species in the EFT with masses below such UV cut-off. 
	
	It is remarkable how the concept of the species scale as a QG cut-off fits very naturally with the ideas of the Distance Conjectures \cite{distance,ADC,review,Irene,Alvaro},
	which claim that the range of validity of any EFT of QG with a fixed UV cut-off is always finite (with some properly defined notion of distance, as e.g. the distance in moduli space in the SDC). This is the case because as one tries to increase this distance, the appearance of infinite towers of states that become light turns out to be inevitable.\footnote{Note that a finite number of states cannot produce such a dramatic breakdown of an EFT.}  As they become light, the states in these infinite towers start to contribute significantly to $N_{\text{species}}$  and from eq. \eqref{species} it can be easily seen that the QG cut-off would then tend to zero.	
	
	In fact, in String Theory examples  there are very often multiple towers which become light and fall below the species scale, so that it is important to analyse which are the 
	relevant species in these cases.  It is also not entirely clear which is the dominant tower in these set-ups and to which tower e.g. the AdS distance conjecture refers to.
	In what follows we make the first steps in clarifying these issues and show how to identify the towers which lie below the species scale and  under which conditions the number of species is given by the 
	product of the maximum number of species in each tower.

	Consider first  the case of only one tower of particles becoming light, with some characteristic mass scale $M_{\text{tower}}$ and spectrum given by
	\begin{equation}
		M_n\, = \, n^{1/p}\, M_{\text{tower}} \, , 
	\end{equation}
	with $n$ some positive integer that labels the step of the state within the tower, and $p$ some finite positive number that characterizes the spacing between the different steps (i.e. it counts the number of towers of particles with identical mass gap, that is e.g. $p=1$ for a single KK tower and $p=2$ for two KK towers with the same mass scale). 
	The connection between the mass of the lightest  particle in the tower and the UV scale is given by an expression of the form 
	\beq
	\Lambda_{\text{UV}}^p \, =\, N \ M_{\text{tower}}^p \,  \longrightarrow 	N^{\frac{D-2+p}{p(D-2)}}\, = \, \dfrac{M_D}{M_{\text{tower}}}\, ,
	\label{bothps}
	\eeq
	where $N$ is the number of species of the tower that enter below $\Lambda_{\text{UV}}$, that is, the maximum value of $n$ for which the states remain within the EFT.
	
	In the presence of more than one (infinite) tower becoming light along the same singular limit, which is the generic case in string compactifications (see e.g. \cite{torres}), it is typically argued that only the tower with lightest mass scale will contribute to $N_{\text{species}}$, and therefore the UV cut-off will only be effectively sensible to it. However, as we will show in the following, this is not always the case and when certain conditions are met it is crucial to include information about heavier towers as they still enter the EFT from this perspective. Let us illustrate this with an example including two towers, with mass spectra given by
	\begin{equation}
		M_{n_1} \, = \, n_1^{1/p_1}\, M_{\text{tower},1} \, , \qquad M_{n_2}\, = \, n_2^{1/p_2}\, M_{\text{tower},2} \, .
	\end{equation}
	
	In a situation of this kind, it is not only the two towers that appear independently, but typically also states with both charges can appear, such that $M_{n_1, n_2}^2\, =  n_1^{2/p_1}\, M_{\text{tower},1}^2 + n_2^{2/p_2}\, M_{\text{tower},2}^2$. This is the case for e.g. two KK towers, or a KK tower and the tower of oscillator modes associated to the fundamental string.\footnote{Here we are making an important assumption about the masses of states charged under both towers, but as long as it is such that it is smaller than the sum of the constituents (i.e. it is stable) similar arguments can be applied.} 
	We would like to compute 	in this situation both the QG cut-off and the species scale associated to such more complex mixed spectra.
	
	\begin{figure}[tb]
		
		\begin{center}
			\subfigure[]{	
				\includegraphics[scale=0.193]{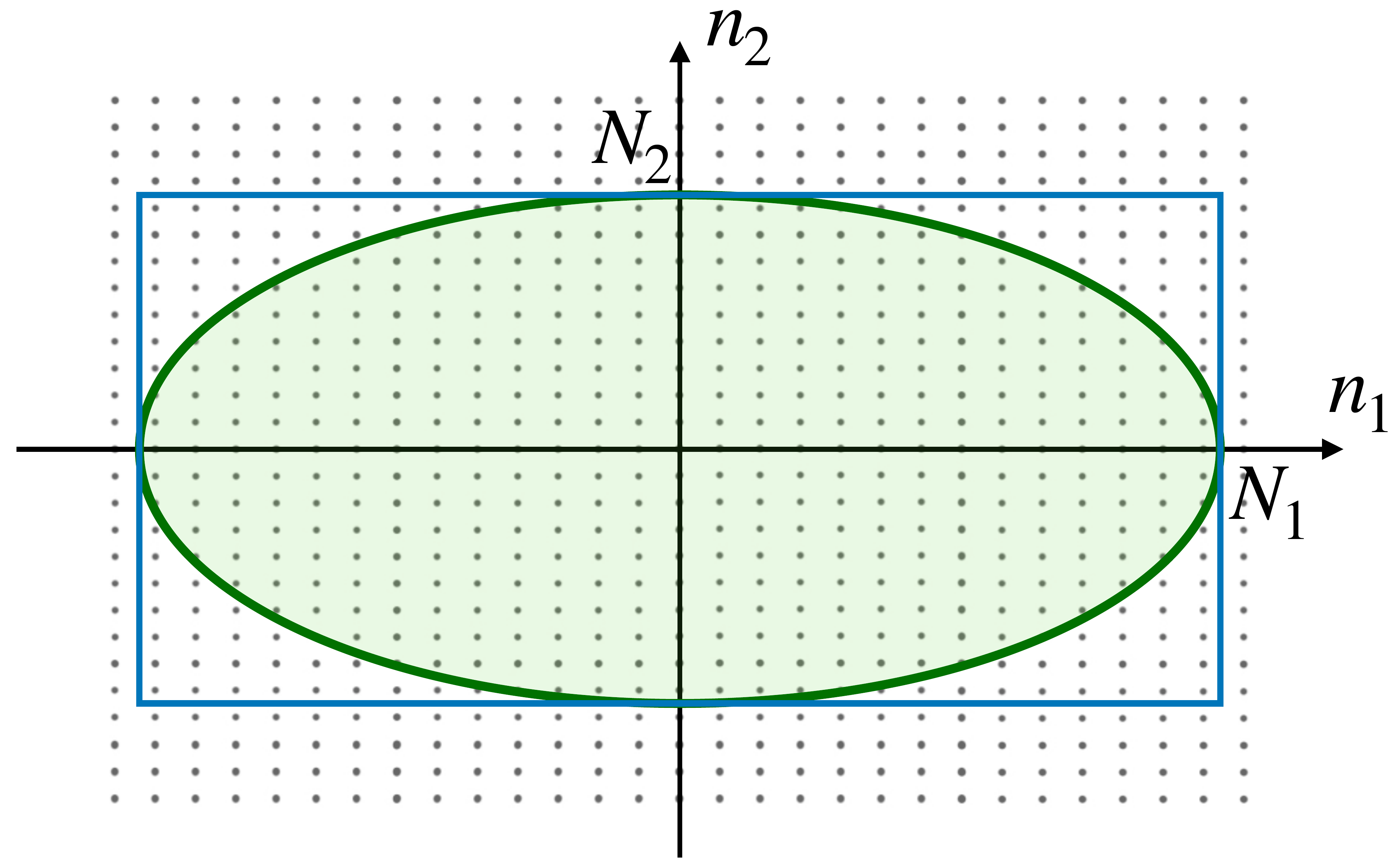}
				\label{fig:2KKtowers}
			}
			\subfigure[]{
				\includegraphics[scale=0.193]{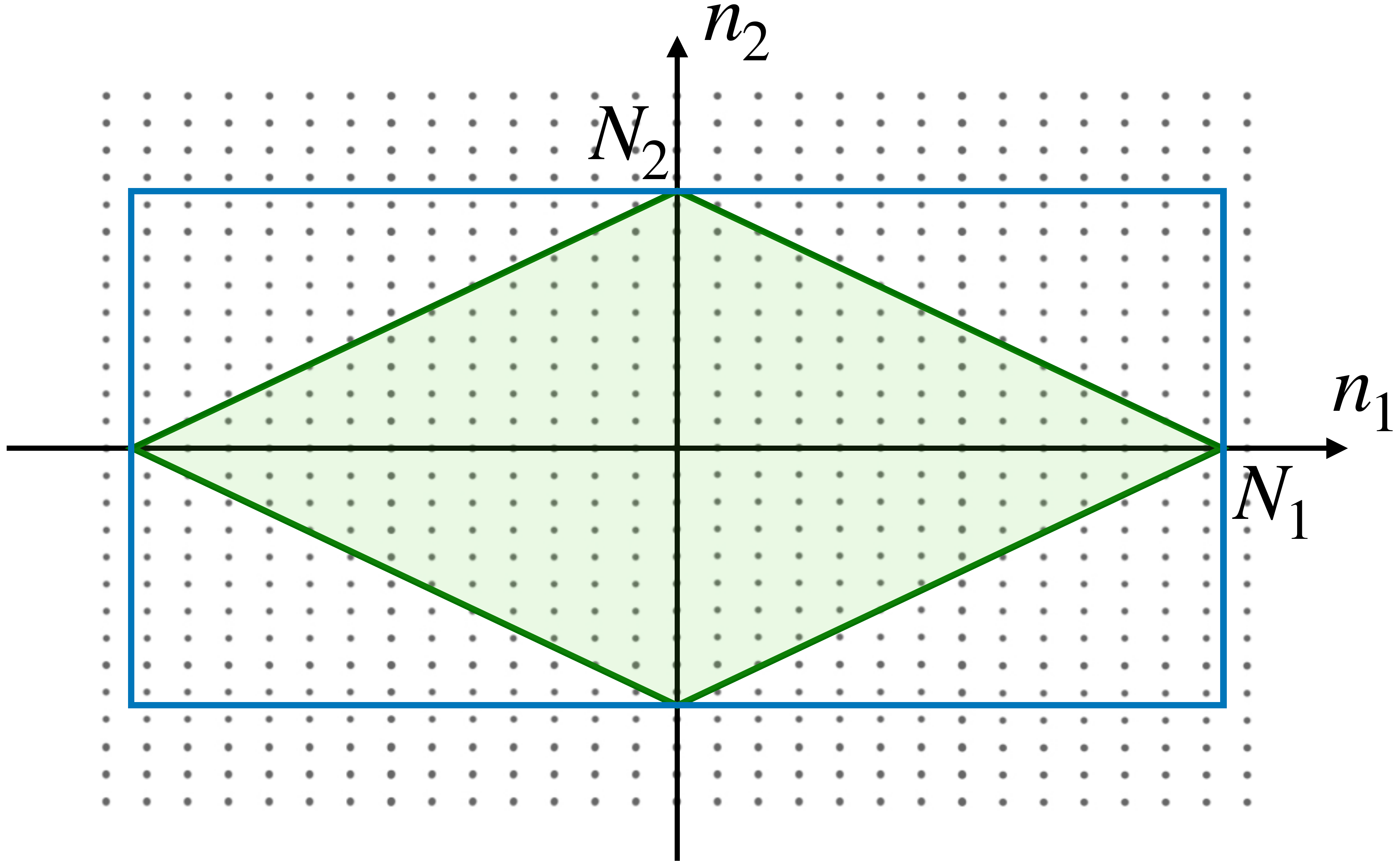}
				\label{fig:2doubletowers}
			}
			\caption{Lattice of states for a light tower with two different mass scales. The states in each tower are labeled by the (positive) integers $(n_1, n_2)$. {\bf (a)} Shows the case of two single KK towers (i.e. $p_1=p_2=1$), whereas {\bf (b)} displays the situation in the presence of two double KK towers (i.e. $p_1=p_2=2$). The green line represents the limiting region where $M=\Lambda_{\text{UV}}$, and its interior (shaded green region) contains all the states that lie below the cut-off and enter the EFT. The maximum excitation numbers $N_1$ and $N_2$ are given by the points where the limiting region cuts the axes. The blue rectangular contour shows the region we are using to approximate the states below the cut-off and it can be seen that for large number of species it differs from the green shaded region only by an order one factor.}
			\label{fig:EFTregions}
		\end{center}
	\end{figure}
	
	Let us consider first the case in which both towers have the same density of states, i.e. $p_1=p_2$, as in the case of e.g. two KK towers along different compact directions. Consider also the case where both mass scales are not parametrically decoupled, such that $M_{\text{tower},2}=c M_{\text{tower},1}$, with $c$ some non-divergent constant. In the limit in which $N_{\text{species}}\to \infty$, the maximal excitation number entering below the species scale will then be of the same order for both towers (exactly equal when $c=1$), such that one is forced to consider states with `mixed charges' (in string compactifications this situation would be relevant for instance when several towers scale equally with the moduli in the strict asymptotic limit).  When there is no parametric decoupling between the characteristic scales of the towers, and given that the maximal occupation numbers, $N_{1,2}$, are both very large, one can approximate the relevant number of lattice sites $(n_1,n_2)$ in the space of allowed charges by integrating the area enclosed by the curve defined by those states with mass equal to $\Lambda_{\text{UV}}$, which is shown in fig. \ref{fig:EFTregions} for two particular cases of KK towers with different structures.
	Moreover, as long as the individual $p_i$ are not parametrically large, as is the case for KK-like towers of particles, this area will only differ by an order one factor with respect to the parallelogram defined by the maximal charges shown also in fig. \ref{fig:EFTregions}.\footnote{Note that this approximation can also be generalized to more towers, but it gets less and less accurate as the number of towers is increased.} 
	One can therefore approximate the total number of species (up to $\mathcal{O}(1)$ factors) by simply the product of the maximal excitation numbers $N_{\text{species}} \sim N_1 N_2$, which yields\footnote{We are ignoring here discrete degrees of freedom like spin, which changes our discussion only by factors of order one.}
	\begin{equation}
		\Lambda_{\text{UV}}^{p_1}\, = \, N_1 \, M_{\text{tower},1}^{p_1}  \, , \qquad \Lambda_{\text{UV}}^{p_2}\, = \, N_2 \,  M_{\text{tower},2}^{p_2} \, , \qquad \Lambda_{\text{UV}}\, \simeq\, \dfrac{M_D}{(N_1 N_2)^{\frac{1}{D-2}}}\, .
	\end{equation}
	
	Let us finally explore the case in which the different towers become light at different rates. Consider for definiteness, the case of two such KK towers, with $M_{\text{tower},2}=c M_{\text{tower},1}$ but now $c$ is allowed to diverge. According to our intuition, the towers seem to present parametric decoupling, so that in principle one can always find some sufficiently high value of $c$ for which as many excitations of the lightest tower (in our conventions the one associated to $M_{\text{tower},1}$) can fit below the first excited state of the heavier one. Thus, naively one would say that it is enough to consider states of the form $(n_1,0)$, with $n_1 \leq N_{1}$ to saturate the total number of species below $\Lambda_{\text{QG}}$ when we are deep inside this regime. Notice that for such argument to work it is crucial that our spectrum is indeed discrete. The only way in which the previous discussion could be wrong and hence fail to compute both $N_{\text{species}}$ and $\Lambda_{\text{QG}}$, would be that when we try to compute the maximum excitation number associated to the second tower that lies below such $\Lambda_{\text{QG}}$ (computed taking into account states charged \textit{only} under the first tower), such $N_{2}$ turns out to be indeed greater than one. That would mean that one cannot just forget about the mixed spectrum of the combined tower, since the states of the lightest one are not enough to saturate the UV cut-off. To get a flavour of what this is telling us, let us simply think in terms of simple KK towers getting light at different rates. Since in this case we know that the species scale computed via the lightest KK tower is essentially the Planck scale of the higher dimensional theory in which the associated cycle has been effectively decompactified, the fact that the second KK scale could still be parametrically below such scale means that its associated tower would be still asymptotically light when measured in higher dimensional Planck units. Thus, this would point towards the necessity of a further decompactification of the associated cycle in order to `resolve' the (infinite distance) singularity in QG. 
	Hence, despite the fact that there is parametric decoupling between the towers, the heavier one could still be light enough to preclude a hierarchy of the form $M_{\text{tower},2} \geq \Lambda_{\text{QG}}\geq M_{\text{tower},1}$. Notice that this does not contradict our initial intuition in the sense that even though in the limit in which $\Lambda_{\text{QG}} \to 0$ one would have $N_{1} N_{2} \to \infty$, still one can check that $N_{1}/N_{2} \to \infty$, such that the lightest tower clearly provides infinitely many more excitations below the cut-off scale than the heavier one. However, still an arbitrarily high number of states from the second tower may enter the EFT. In any event, since we would have to consider now an infinitely large number of points in our lattice of charged states in order to compute both $N_{\text{species}}$ and $\Lambda_{\text{QG}}$, by the same considerations of the previous case of not decoupled towers, one concludes that $N_{\text{species}} \sim N_{1, \text{max}} N_{2, \text{max}}$ up to order one factors. Notice also that this argument can be generalized to the case in which the subleading tower presents a denser spectrum, such as e.g. a double KK tower. Nevertheless, in this case it is not always the case that the tower with lightest mass scale provides infinitely more states than the rest, since the higher density of the heavier modes can indeed compensate and even be dominant in the limit.
	
	Apart from being an interesting refinement of the computation of $N_{\text{species}}$ by itself, the previous argument provides a systematic way to determine when one (or more) towers must be considered  to count the number of species relevant for QG. Moreover, as we will see,  this has important  consequences when analyzing specific String Theory set-ups. Before considering the most general case, let us remark that towers coming from the excitation modes of a critical string becoming tensionless (as e.g. the ones considered in the context of the \textit{Emergent String Conjecture} (ESC) \cite{Lee:2019wij}) can be effectively incorporated into the above reasoning by assigning to them the value $p\rightarrow \infty$. This is due to the fact that, apart from the denser mass spectrum that the stringy tower presents ($m_n  \sim \sqrt{n}\, M_s$ with $n$ being the excitation level of the string), there is the extra degeneracy at each mass level that makes the number of states grow much faster (essentially as an exponential function of $\sqrt{n}$ \cite{Green:2012oqa}) than any finite combination of KK-like spectrum, precisely as if $p\rightarrow \infty$ in the relevant formulae above (as e.g. eq. \eqref{bothps}). 
	
	For an arbitrary number of towers becoming light, the algorithm to compute the species scale and the maximum excitation number of each tower would go as follows:
	\begin{itemize}
		\item[1)]{Take the tower with lightest mass scale in the limit, verifying $M_{n_1} \, = \, n_1^{1/p_1}\, M_{\text{tower},1}$ and compute its associated species scale and number of states $N_{1}$ via eq. \eqref{species}.}
		
		\item[2)]{Compare such $\Lambda_{\text{QG}, 1}$ with the next leading tower, which satisfies $M_{n_2} \, = \, n_2^{1/p_2}\, M_{\text{tower},2}$. Specifically, compute the maximum excitation number of this tower falling below the cut-off scale via the relation $N_{2}=(\Lambda_{\text{QG},1}/M_{\text{tower},2})^{p_2}$. If $N_{2} \leq 1$, then the computation is consistent and the first tower saturates alone the total number of species, i.e. $\Lambda_{\text{QG},(1)}=\Lambda_{\text{QG}}$ and $N_{1}\sim N_{\text{species}}$.}
		
		\item[3)]{If $N_{2} \gg 1$ then the second tower cannot be ignored. One has then $N_{\text{species}} \sim N_{1} N_{2}$ and  an \textit{effective} tower can be defined with mass scale $M_{\text{tower},(2)}^{p_1+p_2}=M_{\text{tower},1}^{p_1} M_{\text{tower},2}^{p_2}$ and density $p_{(2)}= p_1+p_2$ by which one can directly determine $\Lambda_{\text{QG}, (2)}$ via eqs. \eqref{species} and \eqref{bothps}. If there are no more asymptotically massless towers the algorithm terminates here.}
		
		\item[4)]{If there is  further subleading towers one should go back to steps 2) and 3) and iterate the process.}
	\end{itemize}
	
	The final result is then an \emph{effective} tower counting the relevant number of species that lie below the UV cut-off in the limit, with mass scale and density 
	\beq
	M_{\text{tower}} \ =\ (M_1^{p_1} M_2^{p_2} \ldots M_k^{p_k})^{1/\sum_ip_i} \, \qquad p_{\text{eff}}=\sum_i p_i
	\label{centro}
	\eeq 
	given by the (geometric) centre of mass of the constituent towers $M_{\text{tower},i}^{p_i}=M_i^{p_i}$, and the sum of the densities of such towers, respectively. 
	
	Notice that one can also say more about the minimum decoupling that must exist between two `adjacent' light towers in order for the excitation modes of the heavier one to be above the cut-off (in the general case one should think of the leading tower as the effective tower obtained through step 3) of the algorithm above). It can be seen by eq. \eqref{species} that for such adjacent $M_{\text{tower},(i)}$, $M_{\text{tower},i+1}$ with densities $p_{(i)}, p_{i+1}$, such that $M_{\text{tower},(i)} < M_{\text{tower},i+1}$, they must verify the relation 
	\begin{equation}
		M_{\text{tower},(i)} \leq M_{\text{tower},i+1}^{\frac{D-2+p_{(i)}}{p_{(i)}}}\ ,
	\end{equation}
	for the second tower not to contribute to $N_{\text{species}}$. Let us remark that that the main result of this section is the fact that, in the presence of several towers becoming light, it is in general not enough to consider the one with lightest mass scale, as depending on the particular structure of the heavier ones, there can still be an arbitrary high number of them that may lie below the EFT cut-off. Furthermore, note that the above analysis also implies that as soon as the mass scale of a light (critical) string lies below the species scale, so that the corresponding tower (i.e. $p\rightarrow \infty$) has to be included, it automatically saturates the QG cutoff and yields $\Lambda_{\text{UV}}\sim M_s$. This is in agreement with the idea that the species scale captures information about the most fundamental QG scale in the theory, which in this case would be the string scale.
	
	As the analysis above shows, the formulation of the Distance Conjectures  may then suffer from some ambiguity, since one needs to specify to  which tower the conjecture refers. At this point, one could be tempted to propose that it should refer to the tower with lightest mass scale, but as we have seen that is not always the one that provides a higher number of states below the UV cut-off. Another option would be the denser tower, but one could then loose an enormous amount of modes that are lighter than the first mode in the denser tower. Moreover, making an statement like this would implicitly require  some knowledge of the UV completion (i.e. it is a KK tower, a stringy tower, a combination of them, etc.). However, from the point of view of EFT observers,  a tower of states with an intricate structure would become light, where some states might be separated by some  gap and some other states by some different gap. To be precise, a more complicated tower (coming from e.g. a combination of different light towers) as the ones we have studied in this section (see also e.g. fig. \ref{torres3b}), cannot be described only by a mass scale, as in the Distance Conjectures, but one also needs information about the different steps in the tower. From this angle, it makes sense to consider an effective tower which includes all the $N$ states below the species scale and with effective mass, $M_{\text{tower}}$, and effective density parameter $p_{\text{eff}}$, which can be calculated if we know the origin of the tower by means of  eq. \eqref{centro} and $N$ as explained above.  As an example, it turns out that the DGKT-CFI type of models \cite{DGKT,CFI} belong to this class in which several towers appear below the species scale, as discussed at the end of section \ref{sec:ADC}.

	\section{Holography and IR/UV connection}
	\label{sec:holography}

	The argument for a connection between the UV and IR scales is based on the Covariant Entropy Bound applied to a spherical surface. In particular, it has been conjectured that the entropy of a region of space is bounded by the entropy that can be stored in a black hole of the same size \cite{Bekenstein1,Bekenstein2,Bekenstein3,Bekenstein4,Hawking1,Hawking2,tHooft,Susskind}.\footnote{To be precise, in the presence of gravity it is the entropy of the corresponding light-sheet that can be bounded by the area around it, but we restrict here to cases where the \textit{spacelike projection theorem} can be applied and we can therefore bound the entropy within a spacelike region enclosed by the same area, as we will justify later (see \cite{Bousso:1999xy,Bousso:1999cb}).} In a non-gravitational field theory the entropy $S$ is, by construction, an extensive quantity that grows with the volume over which it is defined or restricted. Thus, if we have an EFT in flat space restricted to a (spherical) box of physical size $L$, the entropy of that (sub)system will grow as $L^{(D-1)}$. On the other hand, as argued originally by Susskind in \cite{Susskind}, in the presence of QG the entropy $S_{\text{QG}}$ should grow at most like the surface, namely $S_{\text{max}} \sim L^{(D-2)}$ in Minkowski. Hence, one would naively expect the field theoretical computation of the entropy $S$ to be valid if the box size is restricted in such a way that the entropy is not larger than the covariant entropy bound, i.e. $S\leq A/4$, where $A$ denotes the area of the spherical surface in units in which the Planck length $\ell_P = 1$.  Notice that in the limit in which we decouple gravity, i.e. $G_N \to 0$ (or equivalently $M_D \to \infty$) the entropy bound is trivially satisfied (as befits any meaningful Swampland or QG constraint), since the Bekenstein-Hawking (BH) entropy becomes infinite in that case, so that there is essentially no upper bound for local QFT entropy. 
	
	\begin{figure}[tb]
		\begin{center}
			\includegraphics[scale=0.20]{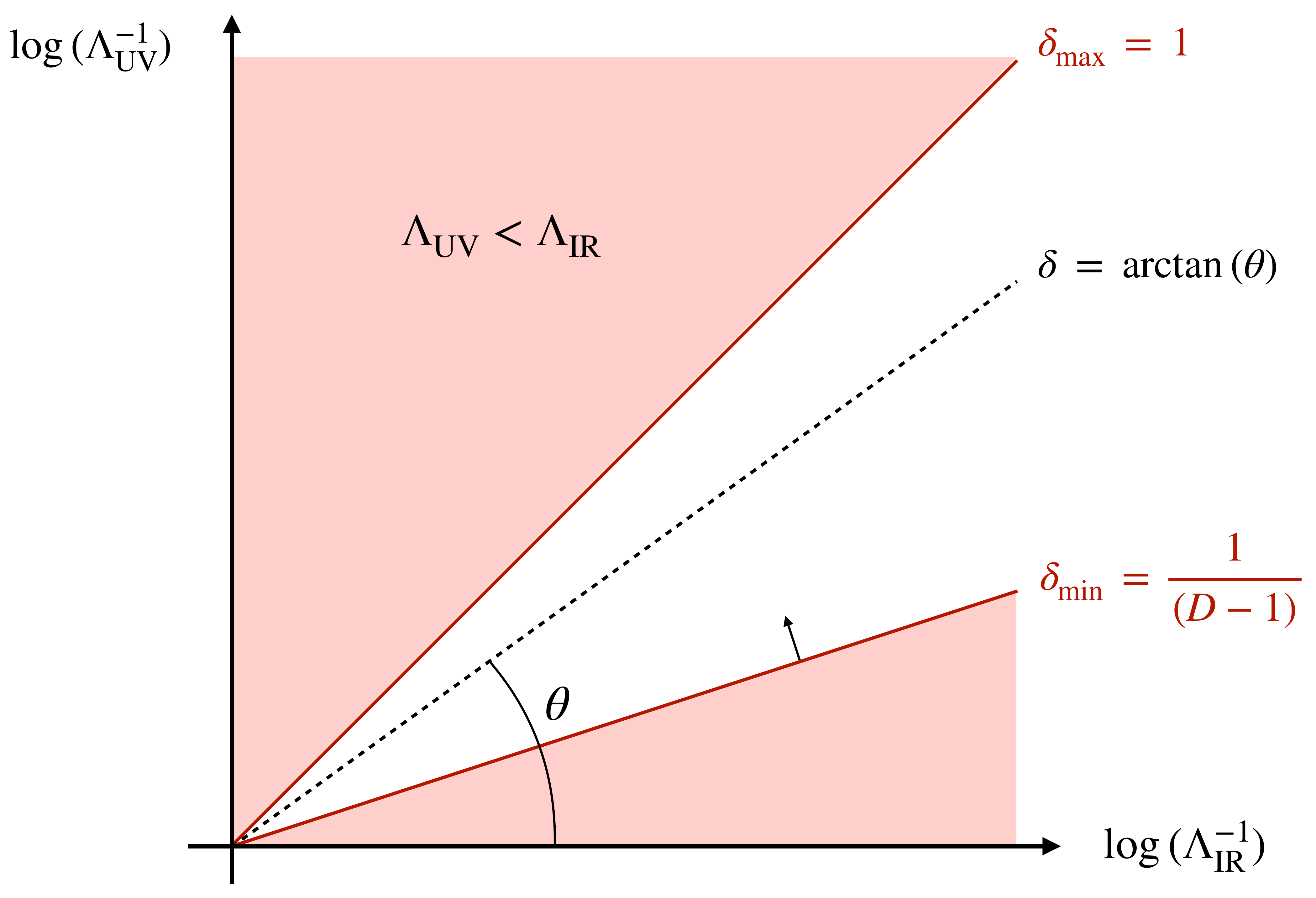}
			\caption{The UV and IR cut-offs in log scale are related linearly. The BH entropy bound implies that there is a minimum slope $\delta_{\text{min}}= 1/(D-1)$.}
			\label{UVIR}
		\end{center}
	\end{figure}

	Let us apply then this general idea to an EFT with UV cut-off $\Lambda_{\text{UV}}$, restricted to a spherical region of size $L$ in $\text{D}\geq 4$ dimensions \cite{CKN,BD}. The (maximal) field theoretical entropy, ignoring possible gravitational corrections due to the weakness of gravity at low energies, then satisfies $S\sim L^{D-1}\Lambda_{\text{UV}}^{D-1}$. On the other hand, the BH entropy goes like $S_{\BH}\sim M_D^{(D-2)} L^{(D-2)}$ instead. Hence insisting in the entropy to remain below the holographic upper bound yields (see fig.\ref{UVIR})
	\beq
	S\leq S_{\BH} \ \longrightarrow \       (L^{D-1}\Lambda_{\text{UV}}^{D-1})\lesssim  M_D^{(D-2)} L^{(D-2)} \longrightarrow  \Lambda_{\text{UV}} \ \lesssim \ (\Lambda_{\text{IR}})^{\frac {1}{ (D-1)} } \ M_D^{(D-2)/(D-1)}\, ,
	\label{eq:BHbound}
	\eeq
	where we have defined $\Lambda_{\text{IR}}=1/L$.
	This expression, which is of the form anticipated in eq. \eqref{general}, explicitly shows that if one decreases the IR cut-off (i.e. one enlarges the IR length scale and hence the size of the box under consideration), the UV cut-off should be lowered too, presenting a clear example of IR/UV correlation in QG.  Note that it may be understood as a smooth transition from an extensive entropy to an holographic one by considering the cut-off as a function of $L$,  $\Lambda_{\text{UV}}(L)$  with
	\beq
	\label{eq:dS_BH}
	dS_{\BH} \ \sim \ \left (\Lambda_{\text{UV}}(L) \right )^{D-1} d(L^{D-1})\ =\ \left (\Lambda_{\text{UV}}(L) \right )^{D-1}\ (D-1) \ L^{D-2}dL \ .
	\eeq
	With $\Lambda_{\text{UV}}$ approximately constant we would recover an extensive entropy. However if instead $\Lambda_{\text{UV}}^{D-1}\sim 1/L$ the entropy will grow like $L^{D-2}$, as expected on holographic grounds \cite{BD}.
	The existence of this UV-IR cut-off correlation in principle affects  e.g. the perturbative loop calculations in an EFT. Thus in a typical Feynman integral the UV and IR cut-offs are not independent from each other, but correlated instead, leading potentially to measurable corrections  \cite{CKN,BD,CK}. 		
	
	Up to now we have been discussing a constraint on the entropy  of  a theory of a finite number of particles with UV cut-off $\Lambda_{\text{UV}}$   in a box of size $L$.
	It is known that 
	in String Theory-derived EFT examples  infinite towers of states indeed appear when going to points located at infinite distances in moduli space  (see reviews \cite{review,Irene,Alvaro}).
	Such a tower of states has direct impact on the fundamental cut-off scale associated to QG, the species scale, which is lowered as the number of particles below it grows.
	It is also worth mentioning that the states in the tower might not interact among themselves, as happens in the case of towers of BPS particles.
	In the spirit of comparing the crude field theory result with the QG one in order to obtain constraints, we will stick to the computation of the field theory entropy associated to the original dof of the EFT, without the inclusion of the tower states, and include the information about the existence of such a tower only via the identification of the cut-off scale with the species scale.
	Now, there is an explicit connection between the UV species scale and the quantities that define the tower of particles, namely its mass scale, $M_{\text{tower}}$, its density of states, $p$ (denoted $p_{\text{eff}}$ in eq. \eqref{centro}), and the number of species below such UV scale, $N$ (see discussion of sect. \ref{sec:towers}). This relation takes the form
	\beq
	\Lambda_{\text{UV}}^p\ = M_{\text{tower}}^p \ N\, .
	\label{bothps2}
	\eeq
	Combining the above expression with the implicit definition of the species scale given in eq. \eqref{species} yields
	\beq
	\Lambda_{\text{UV}}\ =\ M_D^{(D-2)/(D-2+p)}  M_{\text{tower}}^{p/(D-2+p)}\, ,
	\label{torreUV}
	\eeq
	relating the mass scale of the tower to the UV cut-off. Hence, if we further \textit{assume} the $\Lambda_{\text{UV}}$ appearing in eq. \eqref{eq:BHbound} to be precisely the species scale, we arrive at the following relation
	\beq
	\boxed{
		M_{\text{tower}}\ \lesssim \  \Lambda_{\text{IR}}^{2\alpha_D}\ M_D^{1-2\alpha_D}  } \ \ ,\ \ 
	\boxed{ \alpha_D \ =\ \frac {D-2+p}{2p(D-1)} }
	\label{primcota}
	\eeq
	Notice that this discussion easily translates to the case where there is a fundamental string below the species scale, since for this set-up eq. \eqref{torreUV} would tell us that $\Lambda_{\text{UV}} \sim M_s$ as expected, with an $\alpha_D=1/2(D-1)$, i.e. smaller than in the generic case of several KK-like towers becoming light. 
	
\begin{figure}[tb]
		\begin{center}
			\includegraphics[scale=0.25]{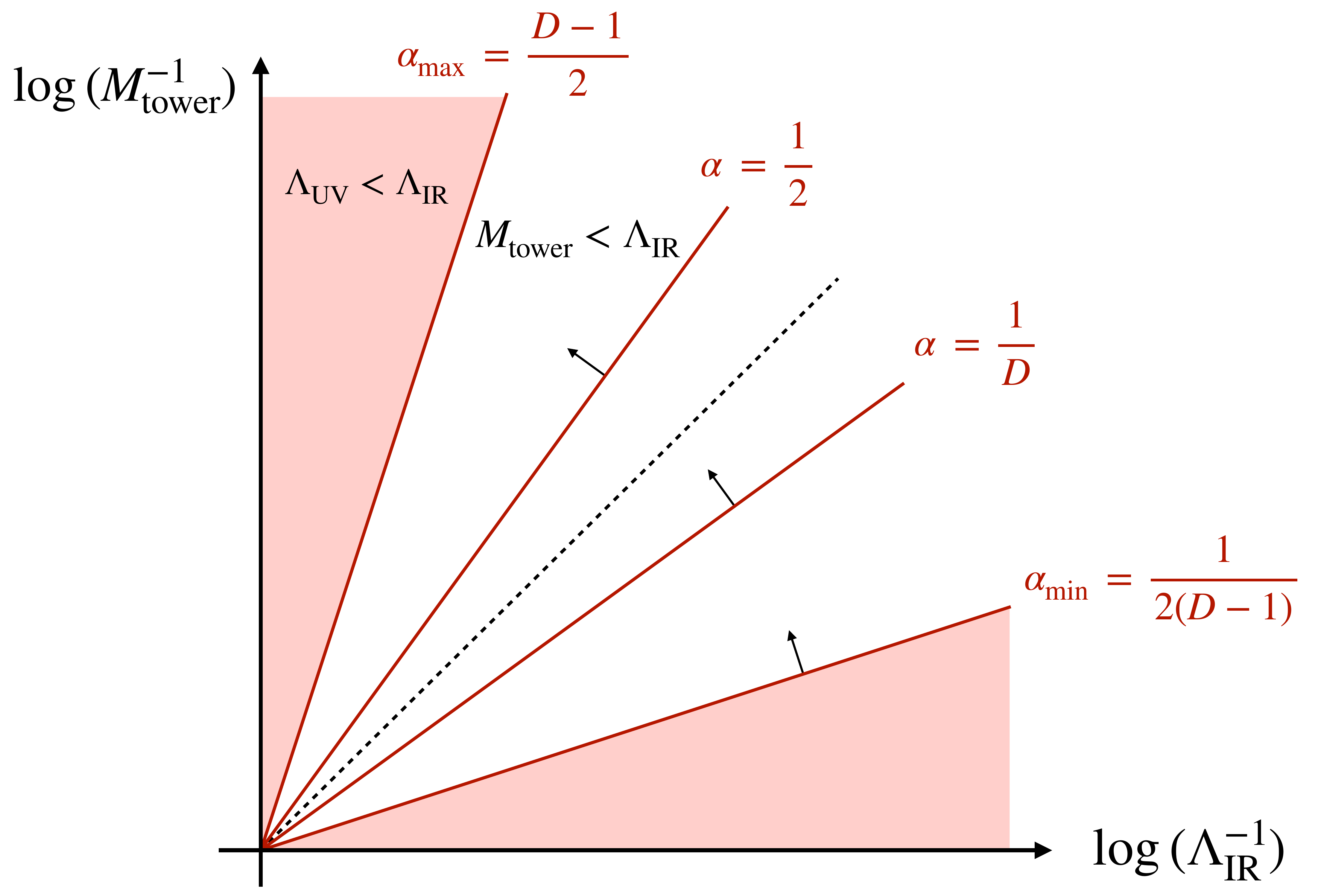}
			\caption{The tower and the IR lengths are linearly related in log scale.  The upper limit for the slope comes from $\Lambda_{\text{UV}}>\Lambda_{\text{IR}}$. The lower bound on the slope 
				comes from the holographic constraint and depends on the number of dimensions D and the type of tower(s) $p$
				as in eq.(\ref{seminal}).  The absolute minimum value $\alpha_{\text{min}}$ is obtained for large $p$ (e.g. in the presence of a string tower) and $\alpha_{\text{max}}$ for $p=1$. However, the CKN bound 
				requires $p\leq D$ and  $\alpha_D\geq 1/D$.}
			\label{MTIR}
		\end{center}
	\end{figure}

	In any event, we thus see that decreasing the IR cut-off the tower of states would also decrease, strengthening the case of a clear IR/UV connection. If one wants this to be a meaningful relation as $\LambdaIR$ is varied (in particular in the limit $\LambdaIR \rightarrow 0$) there must be a relation of the type 
	\begin{equation}
		\Mt \, \sim \,  \LambdaIR^{2 \alpha} \, ,  
	\end{equation}
	with $\alpha\geq \alpha_D$. Moreover, imposing that  $M_{\text{tower}}, \Lambda_{\text{IR}}\leq \Lambda_{\text{UV}}$ one gets a maximum value  for $\alpha$. Altogether this yields 
	\beq
	\label{eq:boundsalpha}
	\alpha_D \ = \ \dfrac {D-2+p}{2p\, (D-1)}\ \leq \alpha \ \leq \ \dfrac {D-2+p}{2p} \ = \ (D-1) \, \alpha_D \, ,
	\eeq
	for any $p$, D. Thus e.g. for $p=1$ in four dimensions, one finds  $3/2\geq \alpha\geq 1/2$. The minimum value for  $\alpha_D$ is found for large $p$, and it gives a general lower bound for $\alpha$ (which could be saturated in the presence of a string tower below the species scale). Similarly, the maximum value for the upper bound is obtained for the minimum value of $p$, which would correspond to the presence of a single KK tower below the species scale. These two considerations combined give the general bounds (see also fig. \ref{MTIR})
	\beq
	\label{eq:totalboundsalpha}
	\dfrac {1}{2(D-1)}\ \leq \alpha \ \leq \ \dfrac {(D-1)}{2} \, .
	\eeq

	Also, let us note that since the species bound is an asymptotic expression, valid when there is a large number $N$ of particles, one expects the constraints derived above to be obeyed in the asymptotic regime of  QG vacua like e.g. string vacua in which a large modulus vev is considered.\newline

	In the above considerations we have included the effect of the tower of states only through its effect on the computation of the species scale.
	This is equivalent to assuming the fields in the tower to be {\it frozen} in the sense that their EFT entropy is not included.
	This gives essentially the right answer if $\Mt\simeq \Lambda_{\text{UV}}$ as in the case of a string tower, in which case the bound on $\alpha$
	is given by $\alpha\geq 1/2(D-1)$.  If that is not the case and one can write $\Lambda_{\text{UV}}=N^{1/p} \Mt$ for some finite integer $N\gg1$, 
	one should in principle include the EFT contribution to the entropy from the fields in the tower.
	If we do so, the bound changes essentially by replacing $\Lambda_{\text{IR}}\rightarrow \Lambda_{\text{IR}}/N$,
	i.e.
	\begin{equation}\label{eq:BetterUVIR}
		\Mt \, \lesssim  \,   \left(\frac {\Lambda_{\text{IR}}}{N}\right)^{2 \alpha}\  \leq \ \Lambda_{\text{IR}}^{2\alpha}\ .
	\end{equation}	
	Thus for fixed $N$, we get a bound of the same form as above and a tower of states will become massless as $\Lambda_{\text{IR}}\rightarrow 0$. However,  the holographic constraint becomes stronger for $N\gg 1$.
	Note that  the weaker bound in eq. \eqref{primcota} only depends on the constant $\alpha_D$,  and the structure of the tower only appears through the parameter $p$.
	On the other hand to make the stronger bound more useful one needs more detailed information about the towers themselves, in particular the number of steps in the tower.
	
	Note that the stronger bound leads to a extreme result for the case of $\alpha=1/2$. Indeed in that case the above condition forces $\Lambda_{\text{IR}}\simeq \Lambda_{\text{UV}}$,
	and then the holographic constraint in eq. \eqref{eq:BHbound} collapses  to  the species bound $\Lambda_{\text{UV}}^{D-2}\lesssim M_D^{D-2}/N$.  For $\alpha >1/2$ one has in general
	$\Lambda_{\text{UV}}\lesssim \Lambda_{\text{IR}}^{2\alpha}/N^{2\alpha -1}$ and a gap between $\Lambda_{\text{IR}}$ and $\Lambda_{\text{UV}}$ opens.
	
	It is illustrative  to see how in the limit in which there is no scale separation and $M_t=M_{KK}\sim \Lambda_{\text{IR}}$ in toroidal compactifications one just recovers the holographic bound in higher dimensions.
	Indeed in the presence of $d$ compact KK towers one has for the entropy and holographic bound
	\beq
	S_{\text{EFT}} \ \sim \ N_0\  \left(\frac {\Lambda_{\text{UV}}}{M_{KK}}\right)^d\ (\Lambda_{\text{UV}}L)^{D-1}\ \lesssim \  L^{D-2} M_D^{D-2} \ .
	\eeq
	with $N_0$ the number of zero modes. In the limit $M_{KK}\simeq \Lambda_{\text{IR}}=1/L$ one gets
	\beq
	S_{\text{EFT}} \ \sim \  N_0 \left( \Lambda_{\text{UV}}L \right)^{D+d-1} \ \lesssim  L^{D+d-2} \ M_D^{D+d-2}
	\eeq
	where we have rewritten the D-dimensional Planck scale in terms of the (D+d)-dimensional one.  One thus recovers as expected 
	the higher dimensional holographic constraint.
	
	\subsection*{Gravitational Collapse and the CKN bound}

	It was pointed out by Cohen, Kaplan and Nelson \cite{CKN,CK} that in fact there is another relevant and seemingly independent cut-off  $\lt$ motivated by gravitational stability of the EFT restricted to the finite spatial region.
	In particular, a connection between UV and IR cut-offs was proposed not directly based on holographic arguments\footnote{However one can argue that whenever we reach the point when the entire system collapses into a black hole just fitting inside the box size, the Bousso bound is still obeyed and indeed saturated.} but rather by avoiding microscopic configurations which would lead to trapped surfaces in the classical geometry sourced by their backreaction. That is, by imposing that, in a system of size $L$,  black holes should not form and be part of the EFT. In the remainder of this section we will consider the situation in which the gravitational UV cut-off is given by such a scale, keeping in mind that the general arguments about gravitational collapse used here might be too naive and not apply in all situations, due to e.g. particular properties (like SUSY) and geometry of the vacuum considered. In any event, we will show that similar bounds for the UV/IR mixing, as well as for the specific parameter $\alpha$ introduced above, can be found if this UV cut-off were to apply, instead of the species scale. It is interesting, though, to point out that all the ten-dimensional String Theory solutions available in the literature that are under perturbative control (e.g. the famous type IIB on $AdS_5 \times S^5$ example), which typically means that they can be described in terms of a low-energy supergravity EFT, verify the gravitational stability constraint with the UV cut-off taken to be the fundamental string scale, $M_s$ (beyond which a pure field theory description is clearly not tenable anymore). 
	
	Gravitational stability imposes then that configurations which lie within their Schwarzschild radius in the box should be excluded.  In D dimensions such condition may be written (up to $\mathcal{O}(1)$ factors) as  
	\beq
	\tilde{N} \lt^D  L ^{D-1} \ \leq \ L^{D-3} M_D^{D-2} \  \ ,
	\eeq
	where $\tilde{N}$ denotes the total number of species falling below $\lt$  and thus entering the EFT. If this number of species is dominated by a tower of particle states of increasing mass, characterized by an effective density parameter $\tilde{p}$ and an effective mass $\Mtt$ we have $\lt \, = \, \tilde{N}^{1/\tilde{p}}\ \Mtt$.\footnote{Note that, as opposed to e.g. eqs. \eqref{eq:BHbound}-\eqref{eq:dS_BH}, here we are explicitly considering the contribution from all the $\tilde{N}$ species of the tower that lie below the UV cut-off (in this case $\tilde{\Lambda}$). This is because whereas in the previous case we wanted to compare the field theory entropy associated to the original EFT (without the tower dof) with the QG bound, in this case we want to explicitly consider the contribution of these extra dof to the calculation of the scale of gravitational collapse.} From here one gets
	\beq
	\label{eq:CKNscale}
	\lt^D \  \leq \ \frac{\Lambda_{\text{IR}}^2M_D^{D-2}}{\tilde{N}} \ .
	\eeq
	Notice that this `CKN cut-off' will be always at or below the species scale $\Lambda_{\text{UV}}$, and above $\Lambda_{\text{IR}}$.
	If this bound indeed applies, it means that the EFT breaks down for energies above $\lt$, since any attempt to excite the quantum fields beyond this point is thwarted by gravitational collapse and 
	the above discussion based on the neat holographic entropy bound  would  only  be valid for states and energies below this cut-off.
	
	What is the total (maximum) entropy of the EFT with the restriction of gravitational stability? In this situation, one can safely proceed with the extensive computation, taking into account the number of particle species $\tilde{N}$ entering the field theory regime
	\beq
	S_{\text{EFT}} \sim \tilde{N} (L^{D-1}\lt^{D-1})\lesssim  \tilde{N}^{1/D} A^{(D-1)/D}\, ,
	\label{eq:CKNentropy}
	\eeq
	where in the formula above we have used eq. \eqref{eq:CKNscale} for the scale associated to gravitational collapse. 
	
	We may write now all the bounds on the different scales with respect to  $\Lambda_{\text{IR}}$.
	The summary (in Planck units)  is:
	\beq
	\label{eq:scales}
	\Lambda_{\text{UV}} \lesssim \ \Lambda_{\text{IR}}^{1/D-1}\, ,  \qquad   \tilde{N}^{1/D} \, \lt \  \lesssim \ \Lambda_{\text{IR}}^{2/D} \, 
	\eeq
	It is interesting to study two limiting cases. On the one hand, by considering the situation with the minimum contribution from the species in the tower, namely identifying the cut-off scale with that of the first state of the lightest tower $\lt=\Mtt$ (i.e. $\tilde{N}\sim 1$),  one sees that eq. \eqref{eq:scales} implies $\alpha \geq 1/D$.\footnote{This bound yields $\Mtt \sim \LambdaIR^{2\alpha}$ with $\alpha \geq 1/D$, but note that in this case this is an upper bound on the mass scale of the lightest tower, which always fulfills $\Mtt \leq \Mt$.
		Here $M_{\text{tower}}$ is  the effective  (geometric average) mass scale defined for the tower of states below the species scale, so that we can only bound the exponent related to $\Mt$ when both coincide.} 
	On the other hand, an upper bound for $\alpha$ can be motivated by asking when the evaporation rate of the would-be black holes (which can be estimated from Hawking radiation rate, $t_{\text{ev}} \sim A^{\frac{D-1}{D-2}}/\tilde{N}$) becomes of the same order of the time scale it takes for the system to collapse ($t_{\text{coll}} \sim A^{\frac{1}{D-2}}$), such that black hole formation can be indeed avoided. In order for this to happen, it must be that $\tilde{N} $, the number of field species below $\lt$ is of the order of the area of the enclosing surface (in Planck units), so that $\lt \sim \Lambda_{\text{IR}}$ is verified and the Bousso bound is moreover saturated. This nicely matches our expectations coming from Swampland considerations, since for this extreme situation one can check using the definition of the species scale given in eq. \eqref{species} that the scale of the three relevant cut-offs in our set-up coincide (i.e. $\Lambda_{\text{UV}} \sim \lt \sim \Lambda_{\text{IR}}$), $\tilde{N} \sim N_{\text{species}}$ and hence the field theory ceases to be valid exactly at the point where its entropy (computed either via the gravitational collapse cut-off or the naive extensive result up to $\Lambda_{\text{UV}}$ as in eq. \eqref{eq:BHbound}) saturates the covariant bound. That is, at this limit, the Bousso bound is saturated by the number of species alone, as expected by the fact that they grow as the area in such a limit. In this case, as all the scales coincide, we also have $\Mt \, = \, \Mtt $ and the upper bound for $\alpha$ is the same as that in eqs. \eqref{eq:boundsalpha}-\eqref{eq:totalboundsalpha}. Notice that for the case in which an asymptotically tensionless string becomes relevant, one finds an upper bound for its decoupling exponent with respect to the IR cut-off, $\alpha_s \leq 1/2$, since otherwise both the CKN and the species scale would fall below $\Lambda_{\text{IR}}$. This result is in accordance with eq. \eqref{eq:boundsalpha} when we take the limit $p \to \infty$.
	
	The comments above suggest that the gravitational stability is not actually independent of the holographic considerations with which we started our discussion in this section. Thus one can show that if we compute the EFT entropy using extensivity up to the point where such description is obliged to collapse into a black hole (which we stress once again would also saturate the holographic bound), we obtain a result of the following form
	
	\beq
	\label{eq:entropygravcollapse}
	S_{\text{EFT}} \sim S_{\text{BH}}^{\gamma} \sim A^{\gamma} \qquad \text{with}\ \gamma_{\text{min}} \leq \gamma \leq \gamma_{\text{max}}\ ,
	\eeq
	where $\gamma_{\text{min}}= \frac{D-1}{D}$ and $\gamma_{\text{max}}=1$, and they are indeed related with the two extreme cases discussed in the previous paragraph, respectively. In particular, $\gamma_{\text{min}}$ is obtained from eq. \eqref{eq:CKNentropy} precisely when $\tilde{N}$ is of $\mathcal{O}(1)$.

	There is an important case worth mentioning at this point. As it is well known, many of the (supersymmetric) AdS vacua of the form $AdS_{D} \times S^d$ (with $D+d=10, 11$) arising from ten-dimensional String Theory or eleven-dimensional M--theory present no parametric scale separation, meaning that the scale of the KK tower saturates eq. \eqref{primcota} with $\alpha=1/2$, regardless of the spacetime dimension of the AdS vacuum
	\cite{Tsimpis,Gautason,Gautason2,Blumenhagen,Font,Apruzzi}. This was in fact conjectured to be true \cite{ADC} in a strong version of the 
	Anti-de Sitter Distance Conjecture, which was proposed to be fulfilled in any supersymmetric vacuum. As it was pointed out in \cite{Aharony:1999ti} and first studied in \cite{Freedman:1983na} (in the particular case of four dimensions), this fact may be deeply rooted in the underlying BPS property of the (massive) particles in ten dimensions, a property which may be crucially preserved under KK compactification on spheres. This implies that the frequency $\omega$ of stationary wave (scalar) solutions in $AdS_{D}$ is quantized in integer multiples of the inverse AdS radius $R_{\text{AdS}}$, such that these fields are indeed single-valued (and thus well defined) in the hyperboloid\footnote{Notice that when one refers to $AdS_{D}$ it is understood that one should consider the universal covering space of such hyperboloid, which has no closed timelike curves to start with.} in which AdS space may be globally embedded within $\mathbb{R}^{2, D-1}$. Now, given that this seems to be a special feature of towers of particles which enjoy the BPS property, one may argue that since such species are thus non-interacting, it may be justified to ignore the gravitational collapse cut-off for this case, such that the computation performed in eq. \eqref{eq:BHbound} could be valid up to the species scale associated to such KK tower. It is interesting that, for the case of $p=1$, our prediction for the $\alpha$ exponent in eq. \eqref{primcota} is precisely $\alpha=1/2$, regardless of the spacetime dimension of our AdS vacuum. From this perspective, this special value (which is indeed allowed by the general bounds eq. \eqref{eq:totalboundsalpha}) would be then just a consequence of having a \emph{single} BPS KK tower, and not a general lower bound as proposed in \cite{ADC}, at least from this crude holographic  perspective. However, one should notice that even though it appears to be the case that in these sphere reductions we have just one simple KK tower, the degeneracy associated to the internal $\text{SO}($d+1$)$ isometry renders the tower to have an effective density of $p_{\text{eff}}=d$, indeed lowering the value for $\alpha_D$.

	\section{ Entropy bounds and the AdS Distance Conjecture}
	\label{sec:ADC}

	Up to now we have not committed ourselves to a specific choice of IR cut-off. However, in the case of QG vacua with curvature, both in AdS and dS, natural choices for this cut-off appear.
	Let us first consider an EFT in AdS space with AdS length $R_{\text{{AdS}}} \sim |V_0|^{-1/2}$, with $V_0$ the cosmological constant/scalar potential evaluated at the minimum in Planck units. 
	In general curved spacetimes, the right way to formulate an area bound for the entropy is through the \emph{Covariant Entropy Bound} \cite{Bousso:1999xy}, which takes the form $S\lesssim A M_D^{(D-2)}$. However, two things need to be taken into account before directly aplying this bound as we did for the Minkowski case. 
	
	First, the Covariant Entropy Bound applies in general to the entropy that corresponds to a \emph{light-sheet} (bounded by a surface whose area enters the bound), which is a null hypersurface instead of a spacelike volume (see \cite{Bousso:1999xy,Bousso:1999cb} for more details on the construction). However, under certain assumptions, one can extend such entropy bound to a spacelike region bounded by the same surface.\footnote{The key non-trivial condition is that the full spacelike region be contained in the causal past of the future directed \emph{light-sheet}. More formally, such a light-sheet must be \emph{complete}.} This is indeed the case for spherical regions of constant proper radius in AdS, so we will consider those as our \emph{boxes} and bound the entropy in the interior of the sphere by its area.
	
	The second important difference with respect to the Minkowski case above is the fact that the area and volume of the corresponding regions must be calculated with the background AdS metric. Given a fixed value of $V_0$, it can be seen that for a sphere of proper radius $L\gg R_{\text{{AdS}}}$ its area and volume scale in the same way, giving the scaling required by the Covariant Entropy Bound, but preventing the obtention of any non-trivial constraints. This is the reason why we will not consider those spheres with very large radii any further. On the other hand, one could consider spheres with very small radius $L\ll R_{\text{{AdS}}}$, but that would just take us back effectively to the Minkowski case, where the curvature of spacetime is negligible and we do not have any hope of getting any non-trivial bound including the cosmological constant. Instead, we will consider as the  natural choice a sphere of radius $L\sim R_{\text{{AdS}}}$, for which the volume and the area scale as $R_{\text{{AdS}}}^3$ and $R_{\text{{AdS}}}^2$ (in four dimensions), respectively. 
	In this case  we will obtain  some non-trivial relations by imposing that the growth of the entropy with the volume be bounded by the area of the boundary (i.e. all the IR/UV relations obtained for asympotically Minkowski spacetimes, like e.g. \eqref{eq:BHbound}, 
	would still hold modulo $\mathcal{O}(1)$ factors) and we can additionally relate the IR cut-off with the value of the cosmological constant via $\Lambda_{\text{IR}}=|V_0|^{1/2}/M_D$.

	In such a case the entropy bound in eq. (\ref{primcota}) implies
	\beq
	M_{\text{tower}}\ \lesssim \ |V_0|^{\alpha_D}\ M_D^{1-D\alpha_D}\ ,
	\label{AdScon}
	\eeq
	with 
	\beq
	\alpha_D \ = \ \frac {(D-2+p)}{2p(D-1)} \ .
	\label{seminal}
	\eeq
	Interestingly, eq. (\ref{AdScon}) has the same content as the ADC of \cite{ADC}. In particular one can see that in the limit of small cosmological constant the tower will become lighter and lighter.
	We have thus  shown that morally the ADC may be understood as a consequence of the holographic principle encoded in the Covariant Entropy Bound. Not only that, the latter gives a precise bound for the exponent for a given number of dimensions and a set of values for $p$ that characterize the relevant towers. 
	Note that for the concrete value $\alpha=1/2$ one has $M_{\text{tower}}\sim 1/R_{\text{AdS}}$ so that there is no meaningful four-dimensional regime with an AdS geometry, there is {\it no scale separation}. That value is the limitting one obtained for any spacetime dimension for $p=1$, i.e. in the case of a single particle tower below the species scale.
	As we said the CKN cut-off gives the lower bound $\alpha\geq 1/D$, if we require the EFT to be valid at least up to the mass scale of the lightest tower. In the cases where this mass scale of the lightest tower can be identified with the effective mass scale for the tower below the species scale (i.e $\Mt \, = \, \Mtt$) this happens for $p=D$. 
	For that value of $\alpha$ one then gets the bound $M_{\text{tower}}\lesssim |V_0|^{1/D}$ for the mass of the lightest tower.
	It is interesting to remark that the values $\alpha=1/2$ and $\alpha=1/D$ were found to be special in the context of the 
	AdS distance conjecture \cite{Rudelius}.

	So far, we have only used general holographic arguments ($S\leq S_{\BH}$) applied to an EFT with infrared cut-off $\Lambda_{\text{IR}}=1/L$. We have derived an upper-bound
	for the UV cut-off, which we identified with the species scale in a theory of QG and considered different possibilities for the composition of the expected degrees of freedom in an UV completion of the theory. 
	We also considered the presence of the lower CKN  cut-off which further constraints the region in which the EFT is trustworthy.
	We would like now  to compare these general results with what actually happens in specific String Theory examples. We have already  seen in section \ref{sec:towers}  that one has to be careful to identify 
	what is the tower or towers of particles that are predicted by the AdS conjecture. The results for the exponent $\alpha$ depend on the detailed structure of the tower that goes down as $V_0\rightarrow 0$. 
	In the simple case of a single tower containing  particles  (and no strings) one identifies $M_{\text{tower}}$ with the lightest, say, KK state. However often in (perturbative) string vacua there are several towers 
	both of particles and a string becoming light below the species scale.
	In \cite{ADC} it was conjectured that any consistent AdS vacuum should have  $\alpha \geq 1/2$, regardless of the spacetime dimension. Indeed, as we said, most of the examples in the literature seem to saturate that bound, 
	with $\alpha=1/2$. From our discussion above, indeed for $p=1$, corresponding to a simple isolated particle tower, like a KK tower, our holographic arguments predict $\alpha\geq 1/2$ for any D, but we could not conclude that the examples in the literature saturate eq. \eqref{AdScon} since for these $p$ is effectively the spatial dimension of the internal sphere. Moreover, all these
	models do not have scale separation between the KK and AdS scales (see fig.{\ref{torres3a}).

		\begin{figure}[tb]
			
			\begin{center}
				\subfigure[]{	
					\includegraphics[scale=0.135]{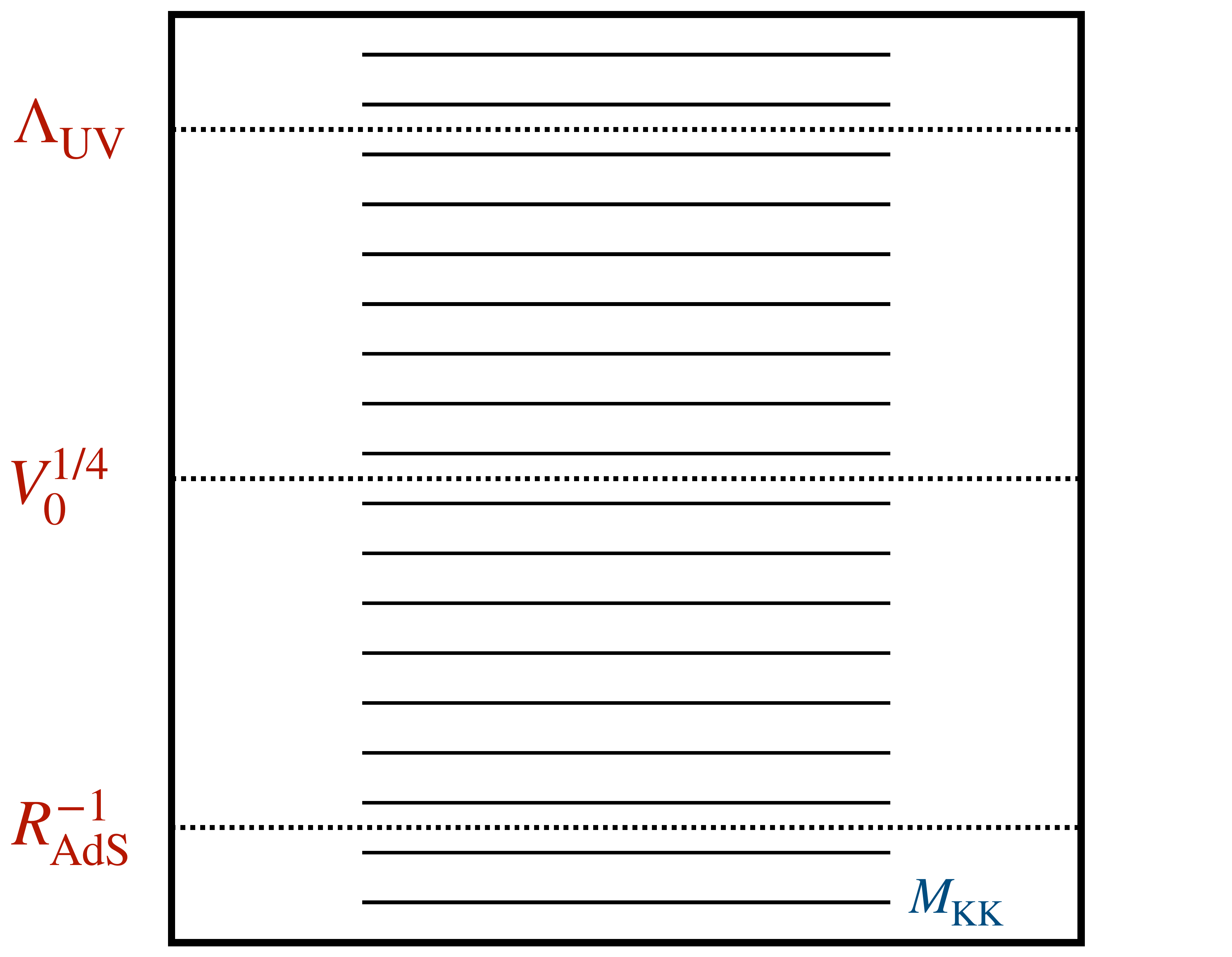}
					\label{torres3a}
				}
				\subfigure[]{
					\includegraphics[scale=0.135]{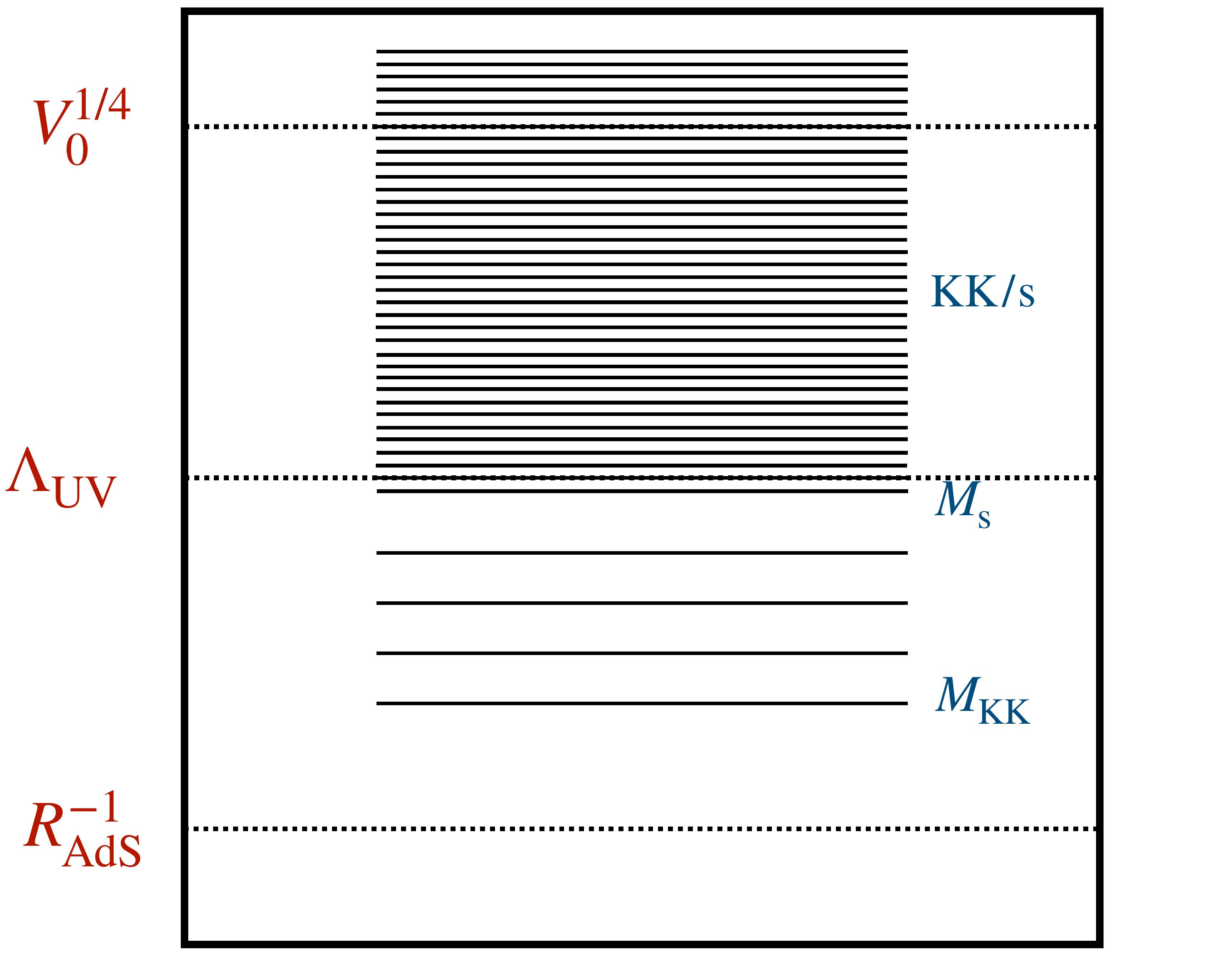}
					\label{torres3b}
				}
				\subfigure[]{
					\includegraphics[scale=0.135]{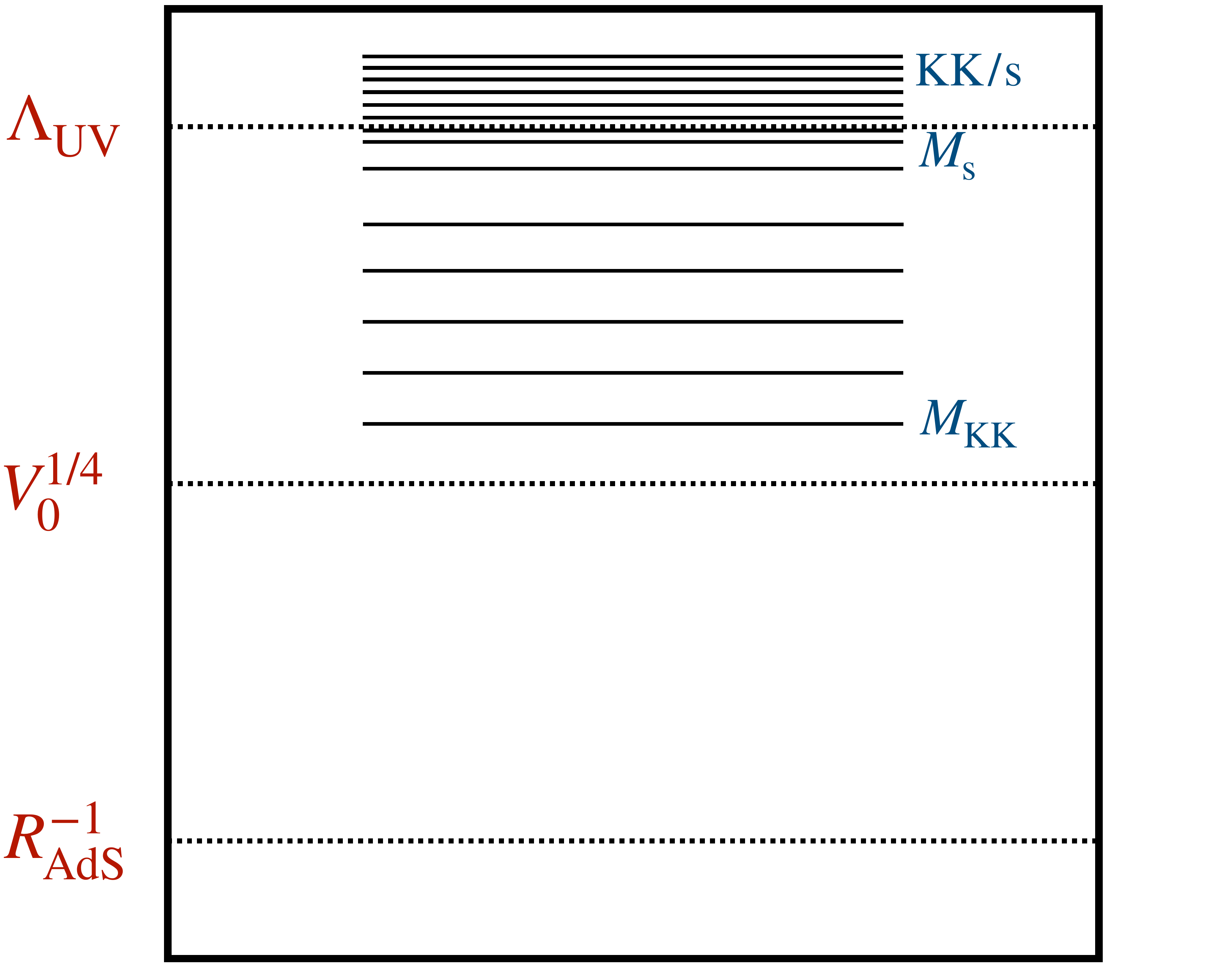}
					\label{torres3c}
				}
				\caption{Structure of towers and scales in four dimensions: {\bf (a)} A theory with a single tower of particles with $\alpha=1/2$. No scale separation. {\bf (b)} The structure of towers in DGKT-CFI models, with two towers of particles and one of strings.
					There is  scale separation, with some part of the tower below  the maximum CKN cut-off $V_0^{1/4}$. {\bf (c)} A theory with the  towers above the CKN bound.}
				\label{torres3}
			\end{center}
		\end{figure}

		There is a class of four-dimensional Type IIA AdS orientifold vacua first constructed (independently) in \cite{DGKT} and \cite{CFI}, which have $\alpha =7/18$, with the tower identified with the KK tower in those models, violating the bound $\alpha\geq 1/2$ proposed in \cite{ADC}. Thus, in light of the above discussion, one may ask whether this class of models is in agreement with our holographic bounds or not. The first important point to realize, following our arguments in section \ref{sec:towers}, is that in these particular examples the structure of towers becoming light below the species scale is more complicated, involving not only three KK towers (corresponding to the three independent 2-cycles of the $T^6/\mathbb{Z}_3^2$ or $T^6/\mathbb{Z}_2^2$ orbifold, hence with $p=2$ each of them) but also a fundamental string tower.
		To see this we refer the reader to fig. 2(b) in \cite{torres}, where the structure of scales in Type IIA models in which the relevant moduli fields scale like $s\sim u \sim t^3$ is shown, which is the case in the DGKT-CFI models. 
		Here $s$ is the four-dimensional dilaton and $t$ is the overall K\"ahler modulus of the particular isotropic case. 
		Hence, one sees that there are in principle four towers in the lightest sector, three KK towers and a string tower, with masses scaling (for fixed $M_4=1$) like $M_{\text{KK}}\sim t^{-7/2}, M_s\sim t^{-3}$, so that the lightest states are the KK species. To see whether we can consider this KK tower in isolation we follow the algorithm in section \ref{sec:towers} to compute the species scale.
		One starts by computing the species scale associated to just  the KK towers to  find that $\Lambda_{\text{KK}}\sim t^{-21/8}$, which would be above the string scale. Hence, since one is forced to take into account the string modes in order to compute the species scale for this set-up, one concludes that it is indeed saturated by the string excitations and thus $\Lambda_{\text{UV}} \sim M_s = M_{\text{tower}}$. The number of species below the cut-off  $\Lambda_{\text{UV}}$ for each tower $N_{\text{KK}},N_s$ is thus computed as follows
		\beq
		\Lambda_{\text{UV}} \ \sim \ M_s \ , \ N_s \ \sim \ \left(\frac{M_4}{M_s}\right )^{2} \ ,\  \ \Lambda_{\text{UV}} \ =\  M_{\text{KK}} N_{\text{KK}} \ .
		\eeq
		This way one can estimate how the number of species of each kind grow with the modulus, $N_s\sim t^{6}$ and $N_{\text{KK}}\sim t^{7/6}$. Thus for large modulus the number of species is dominated by the string tower (contrary to our naive expectations), although the presence of the other kind of states cannot be ignored. So from holography arguments we obtain in this particular example that $\alpha \geq 1/6$ for $M_{\text{tower}}\, \sim \, M_s$. Thus the holographic bound for this structure of towers is (see fig. \ref{torres3b})
		\beq
		M_s\ \lesssim V_0^{1/6}\ .
		\eeq
		On the other hand one can go to the specific results for the spectrum of this class
		of models and find out that the minima sit at $V_0\sim t^{-9}$ and $M_s\sim t^{-3}$ so the bound is indeed obeyed, although not saturated. 
		
		If we however insist in cutting off our field theory at the CKN scale of gravitational collapse, $\lt$, it must be the case that the lightest tower (which happens to be the KK one in this set-up) should verify $M_{\text{KK}} \sim \LambdaIR^{2\alpha}$ with $\alpha \geq 1/4$, as explained in section \ref{sec:holography}. Notice that this is precisely satisfied in these vacua since $\alpha = 7/18$ for the KK towers, but not saturated, meaning that the true CKN scale may be lower than $V_0^{1/4}$ and it may take into account even species coming from the string modes (which also lie below the maximum CKN cut-off $V_0^{1/4}$).

		There is some debate about whether the DGKT-CFI models, which feature scale separation (in the sense that one can keep $M_{\text{tower}}\gg  V_0^{1/2}/M_4$) have some inconsistency (see e.g. \cite{Acharya,Petrini,Saracco,Sethi,Marchesano,Junghans,Buratti,Cribiori,DeLuca}). 
		We see that from the point of view of the four-dimensional  EFT the spectrum is consistent with the holographic analysis performed here. There is also the point that we have found that the string scale lies below the (maximum) CKN cut-off $V_0^{1/4}$, so that it may be argued that this may affect  the computation of the effective (flux) moduli potential. This may make difficult the uplifting to a fully ten-dimensional field theory.
		It would be interesting to test eq. \eqref{AdScon} in other AdS vacua in different dimensions $D>4$. To our knowledge, no candidate for a QG vacuum  exhibiting full scale separation in the sense of fig. \ref{torres3c} (i.e. with the mass of the lightest tower above $V_0^{1/D}$) has been found so far.
		
		There are more general situations in which a tower of infinite (asymptotically massless) states is expected to appear. Thus in \cite{gravitino2,gravitino} it was conjectured that in theories with a massive gravitino the limit $m_{3/2}\rightarrow 0$ is at infinite distance, i.e. one has in that limit 
		\beq
		\frac {M_{\text{tower}}}{M_D} \ \lesssim \ \left(\frac {m_{3/2}}{M_D}\right)^\delta \ \ ,\ \  0<\delta \leq 1 \ .
		\label{gravit}
		\eeq
		This {\it gravitino distance conjecture} is identical to the AdS conjecture in (supersymmetric) AdS space with $\delta=2\alpha$, but it applies also to Minkowski and dS spaces (see also \cite{Cribiori:2020use,DallAgata:2021nnr} for related works on vanishing gravitino masses and dS critical points). 
		In four dimensions one obtains a lower bound based on the supergravity EFT $\delta\geq 1/3$ \cite{gravitino}, corresponding to $\alpha\geq 1/6$, in agreement with the holographic bounds above (see eq. \eqref{eq:totalboundsalpha}).
		One may consider deriving this gravitino conjecture from similar holographic arguments by choosing $\Lambda_{\text{IR}}=m_{3/2}$. In SUSY AdS this is fully justified since $m_{3/2}=|V_0|^{1/2}$ in Planck units,  but such an identification is far from clear for more general backgrounds.		
		
		Let us end this section by pointing out that the present line of reasoning also works in the inverse direction. If we assume the AdS distance conjecture to hold, with no scale separation ($\alpha=1/2$) for a KK(-like) tower, and the expression for the species scale $\Lambda_{\text{UV}}=1/N^{1/D-2}$, then one re-derives back the initial holographic bound eq. \eqref{eq:BHbound}. Indeed, with $\Lambda_{\text{UV}}=N M_{\text{tower}}$ one gets $\Lambda_{\text{UV}}=M_{\text{tower}}^{1/D-1}$. With no scale separation  we have then $M_{\text{tower}}=V_0^{1/2}=\Lambda_{\text{IR}}$ and indeed one recovers eq. \eqref{eq:BHbound}. Thus in some way the form of the AdS conjecture for models with a KK tower secretly contains that holographic structure.

		\section{Entropy constraints on dS vacua and our universe}
		\label{sec:dS}

		For the case of (quasi-)dS vacua,\footnote{One can think of these as (non-empty) spacetimes that approach de Sitter space asymptotically both in the future and in the past, usually denoted as dS$^{\pm}$.} one can use similar holographic arguments in order to constrain the validity of any EFT defined within it. Thus, since dS space has for any observer sitting at an arbitrary location a \textit{cosmological horizon} defined at $r=R_{\text{dS}} \sim V_0^{-1/2}$ in Schwarzschild coordinates, the entropy associated to such EFT cannot exceed the Gibbons-Hawking bound (which is once again an special instance of the more general covariant entropy bound along with the spacelike projection theorem), i.e. in $\text{D}=4$, $S_{\text{EFT}} \leq S_{\text{GH}}=\pi R_{\text{dS}}^2$, in units in which $G_N=1$
		(see e.g.\cite{BD} for a more complete discussion of holographic bounds in dS space). Thus, since the IR cut-off must be at or below the dS radius $R_{\text{dS}}$, a similar relation than the one in eq. (\ref{primcota}) should hold. 
		Then we would obtain a constraint analogous to the one above in AdS vacua.
		In other words, one also expects for dS vacua that
		\beq
		M_{\text{tower}}\ \lesssim \ V_0^{\alpha_D}\ M_D^{1-D\alpha_D}\ ,
		\label{dScon}
		\eeq
		with a general expression for $\alpha$ as in eq. \eqref{seminal}.	
		Indeed the fact that the AdS distance conjecture should be extensible to the case of dS vacua was stated already in the original formulation of the ADC in \cite{ADC}. There it was conjectured  that in the dS case one would rather have a reversed bound $\alpha\leq 1/2$. An argument in favour of this latter inequality would be the Higuchi bound \cite{Higuchi:1986py,Lust:2019lmq} applied to a possible spin-2 mode in the tower of mass $m_{J=2}$, as typically arises from KK replicas of the graviton. Unitarity in dS vacua imposes that $m_{J=2} \geq \ \Lambda_{\text{cc}} ^{1/2}$. Instead, what we find is that  the holographic arguments discussed here give the same bounds both for dS and AdS vacua. However, we have seen that values $\alpha\leq 1/2$ are in principle allowed by the minimum $\alpha_{\text{min}}=1/D$.
		However, consistency will require that the tower must be more complex than just a single KK tower. Towers of states becoming massless are thus expected when $V_0$ goes to zero both from negative and positive values.

		In the asymptotic regime here considered the dS Swampland conjecture \cite{dS,K} implies a behaviour $V'/V\gtrsim c$, with $c$ a constant of order one in Planck units. 
		Following ideas similar to those in \cite{Andriot} (see also \cite{TCC})  one may relate this constant $c$ to the SDC parameter $\lambda$ and the parameter $\alpha$ here considered.
		Indeed one may  (qualitatively)  write 
		\beq
		M_{\text{tower}}\ \sim \ V_0^\alpha \ \sim e^{-\lambda \Delta} .
		\eeq
		For an exponential behaviour of $V_0$ in terms of a modulus $\phi\sim \Delta$, one expects
		\beq
		c\ =\ \frac {\lambda}{\alpha} \ .
		\eeq
		This is just a generalisation of the conjecture that $c=2\lambda$ put forward in \cite{Andriot} for general $\alpha\not=1/2$. 
		
		We can try to apply the holographic  bounds discussed in this note to the observed universe which seems to live in a (quasi-)dS phase. In particular, if we consider the infrared cut-off to be $\Lambda_{\text{IR}}=V_0^{1/2}/M_4$ as above, the holographic bound eq. \eqref{eq:BHbound} in four dimensions gives rise to 
		\beq 
		\Lambda_{\text{UV}}\ \lesssim \ \Lambda_{\text{IR}}^{1/3}M_4^{2/3} \ =\ V_0^{1/6}\ M_4^{1/3} \ .
		\eeq
		Setting $V_0=(10^{-2}\ \text{eV})^4$ one obtains $\Lambda_{\text{UV}}\lesssim 10^{-2}$ GeV, which is against the experimental fact that we have not observed the
		fundamental scale of QG at such low energies.  This fact was pointed out in the past (see e.g.\cite{Rudelius}) for the case of a KK tower with $\alpha=1/2$.
		However, our statement here is more general in the sense that it is independent of the value of $\alpha$ and applies regardless of the internal structure of towers, as long as the holographic bound eq. \eqref{eq:BHbound} applies and one identifies $\Lambda_{\text{IR}}=V_0^{1/2}$.
		It could be that the universe is not in a dS vacuum but in a runaway dS phase and our bounds need to be
		modified. Certainly this issue deserves further study.\footnote{The fact that the AdS conjecture (as applied also to dS) could give rise to a tower of massless states in our universe was already pointed out in ref. \cite{ADC}.}  
		
		One may consider as an alternative candidate for the IR cut-off the height of the potential
		$\Lambda_{\text{IR}}=V_0^{1/4}$. Then one would get in four dimensions
		\footnote{A physically motivated scale  $L$ would also be the smallest one fitting all the Compton wave-lengths of the SM 
			massive spectrum. In this case $\Lambda_{\text{IR}}\simeq m_{\nu_1}$, with $m_{\nu_1}$ the mass of the lightest neutrino. 
			The latter is  expected to be lighter than $0.1$ eV, probably somewhat lighter. This is numerically of order $V_0^{1/4}$ 
			and hence the scales discussed below also apply to this cut-off.}
		\beq
		\Lambda_{\text{UV}} \ \lesssim \ V_0^{1/12}\ M_4^{2/3} \ ,
		\label{esta}
		\eeq
		leading to $\Lambda_{\text{UV}}\lesssim 10^8$ GeV, which would be  consistent with experimental bounds. 
		Also for the CKN cut-off one finds 
		\beq
		\lt \sim V_0^{1/8}M_4^{1/2}\sim 10^{3.5}\text{GeV}
		\eeq
		just above the LHC energies and hence consistent with experiment.
		For the scale of the associated tower would be		
		\beq
		M_{\text{tower}}\ \lesssim \ V_0^{\alpha /2}\ M_4^{1-2\alpha} \ .
		\label{torrecilla}
		\eeq
		Numerically, for $\alpha=1/4$ one has again $M_{\text{tower}}\simeq 10^{3.5}$ GeV, above LHC scales.
		This would provide for an understanding of the stability of the EW hierarchy.
		For $\alpha=1/2$ one would rather get   $M_{\text{tower}}\sim V_0^{1/4}\sim 10^{-2}$ eV.
		In that case such tower should not couple to the SM directly, since otherwise it would have been already observed.
		It would be interesting to see whether one can somehow motivate this choice of IR cut-off when applying these arguments 
		to the observed universe.

		\section{Final comments}
		\label{sec:conclu}
		
		In this  note we have presented heuristic arguments based on the Covariant Entropy Bound which motivate an explicit correlation between IR and UV cut-offs 
		in an EFT in a box of size $1/\Lambda_{\text{IR}}$. In particular one obtains $\Lambda_{\text{UV}}\lesssim \Lambda_{\text{IR}}^{1/(D-1)}$ in Planck units.
		We argue that in a theory of QG one should then identify the UV cut-off $\Lambda_{\text{UV}}$ with the species scale
		so that  as $\Lambda_{\text{IR}}\rightarrow 0$ towers of species below $\Lambda_{\text{UV}}$ become massless.
		Typically in  e.g  string vacua when  $\Lambda_{\text{UV}}\rightarrow 0$ multiple towers of states (both particles and/or string excitation modes) become light, so that 
		a careful analysis is needed in the general case to identify their structure and density of states. We give a general algorithm to calculate 
		what the species scale is in the case of multiple towers with particles and strings, and obtain a general upper bound on the mass of the effective tower of the form $M_{\text{tower}}\lesssim \Lambda_{\text{IR}}^{2\alpha_D}$, with $\alpha_D$ given by eq. \eqref{primcota}. The latter depends both on the dimension D and on the structure of the tower through the density parameter $p$. Thus we obtain that when the IR cut-off goes to zero, towers of species become light in a very definite pattern. In the case of a single particle (e.g. KK) tower,  $M_{\text{tower}}=M_{\text{KK}}$, but in the case of multiple towers $M_{\text{tower}}$ is rather a geometric mean of the scale of the towers present below the species scale, weighted by their corresponding density, whilst the presence of a critical string is enough to saturate the species bound. These results have the caveat that before the species scale $\Lambda_{\text{UV}}$ is reached, gravitational collapse may appear at a scale $\lt$ smaller than the cut-off. This makes fully reliable only the EFT below $\lt$ which implies further constraints and in particular $\alpha \geq 1/D$ for the lightest species.

		Identifying $L$ with the curvature radius in AdS vacua, $R_{\text{AdS}}$, this expression implies the  AdS Distance conjecture, with an specific lower bound for the $\alpha$ exponent, c.f. eq. \eqref{seminal}. For only one light KK tower one predicts $\alpha\geq 1/2$ in any spacetime dimension, in agreement with the expectations in \cite{ADC}. However for string or multiple KK towers one rather finds a general range $\alpha_D\leq \alpha\leq (D-1)\alpha_D$ (c.f. eq. \eqref{eq:boundsalpha}). Thus for $p>1$ values $\alpha<1/2$ may be indeed allowed by entropic arguments.
		Most AdS vacua examples in the literature saturate the Covariant Entropy Bound constraint with $\alpha=1/2$, although the reason behind this could be related to the presence of a BPS KK tower in such supersymmetric set-ups. Concerning the DGKT-CFI class of models \cite{DGKT,CFI}, we find that they feature multiple towers below the species scale (including the fundamental string tower), obeying (but not saturating) the holographic bound.
		However we point out that in this class of models one has $M_{\text{string}}< |V_0|^{1/4}$, which may indicate that one cannot ignore the string tower when
		computing the full (scalar) potential, as customarily done in the literature.
		
		On the other hand, the different asymptotic  Swampland conjectures, namely the Distance, AdS Distance and dS Conjectures are expected to be connected (see e.g.\cite{Andriot,TCC} and references therein). Hence one might expect that their corresponding rate parameters $\lambda$, $\alpha$ and $c$ may also be computed/bounded using analogous arguments based on IR-UV mixing from entropy considerations as done in here. 
		We find here a connection between these three parameters giving $c=\lambda/\alpha$. We also expect that similar arguments to the ones presented in this work can be used to bound the parameter $\lambda$ in the SDC. Also, in the context of (quasi-)dS vacua in QG, it would be interesting to study the relation between the kind of arguments presented here and the Festina Lente bound \cite{FL1, FL2}.
		
		One may consider as well the application of these bounds to the present universe, which seems to be in a dS phase. If one takes as the 
		IR cutoff $\Lambda_{\text{IR}}\sim V_0^{1/2}/M_4$, the resulting UV cut-off  is of order 10 MeV, which seems to be excluded experimentally. 
		If one rather takes as IR cut-off $\Lambda_{\text{IR}}\sim V_0^{1/4}$, one gets $\Lambda_{\text{UV}}\sim 10^8$ GeV with $\lt\sim 10^{3.5}$ GeV, just above the LHC and
		with no apparent contradiction with experiment.  However, a motivation for this second choice is missing.

		The results in this work give additional evidence for a correlation between IR and UV cut-offs in a theory consistent with  Quantum Gravity. The simple power behavior for the cut-off $\Lambda_{\text{UV}}\sim \Lambda_{\text{IR}}^{1/(D-1)}$ 
		or for the tower of states $M_{\text{tower}}\leq \Lambda_{\text{IR}}^{2\alpha_D}$ are consistent with explicit examples of string vacua. 
		In particular, these arguments give an heuristic understanding of the AdS Distance Conjecture providing the possible rationale behind it, and moreover they allow us to give bounds on its relevant parameters. Still the simple argumentation in terms of the Covariant Entropy Bound here used is very 
		preliminary, and questions like the gravitational stability of the EFT definitely require further study. 
		
		 \newpage
		 
		\centerline{\bf Acknowledgments}
		\vspace{.5cm}	
		We are grateful to  F. Marchesano, M. Montero, A. Uranga,  I. Valenzuela and M. Flory for useful discussions. 
		This work is supported  by  the  Spanish  Research  Agency  (Agencia  Espa\~nola  de  Investigaci\'on) through  the  grants  IFT  Centro  de  Excelencia  Severo  Ochoa  SEV-2016-0597, the grant GC2018-095976-B-C21 from MCIU/AEI/FEDER, UE and the grant PA2016-78645-P. The work of A.C. is supported by the Spanish FPI grant No. PRE2019-089790. The work of A.H. is supported by the ERC Consolidator Grant 772408-Stringlandscape.

	\end{document}